\documentclass[reprint,amsmath,amssymb,aps,twocolumn,showkeys,superscriptaddress]{revtex4}
\usepackage{graphicx}
\usepackage{dcolumn}
\usepackage{mathrsfs}
\usepackage{xcolor}
\usepackage{appendix}

\begin{document}

\title{Elastic Microphase Separation Produces Robust Bicontinuous Materials}


\author{Carla Fern{\'a}ndez-Rico}
\affiliation{Department of Materials, ETH Z\"{u}rich, 8093 Zurich, Switzerland}%
\author{Sanjay Schreiber}
\affiliation{Department of Materials, ETH Z\"{u}rich, 8093 Zurich, Switzerland}%
\author{Hamza Oudich}
\affiliation{Department of Mechanical and Process Engineering, ETH Z\"{u}rich, 8092 Zurich, Switzerland}%
\author{Charlotta Lorenz}
\affiliation{Department of Materials, ETH Z\"{u}rich, 8093 Zurich, Switzerland}%
\author{Alba Sicher}
\affiliation{Department of Materials, ETH Z\"{u}rich, 8093 Zurich, Switzerland}%
\author{Tianqi Sai}
\affiliation{Department of Materials, ETH Z\"{u}rich, 8093 Zurich, Switzerland}%
\author{Stefanie Heyden}
\affiliation{Department of Materials, ETH Z\"{u}rich, 8093 Zurich, Switzerland}%
\author{Pietro Carrara}
\affiliation{Department of Mechanical and Process Engineering, ETH Z\"{u}rich, 8092 Zurich, Switzerland}%
\author{Laura De Lorenzis}
\affiliation{Department of Mechanical and Process Engineering, ETH Z\"{u}rich, 8092 Zurich, Switzerland}%
\author{Robert W. Style}
\affiliation{Department of Materials, ETH Z\"{u}rich, 8093 Zurich, Switzerland}%
\author{Eric R. Dufresne}
\affiliation{Department of Materials, ETH Z\"{u}rich, 8093 Zurich, Switzerland}%
\email{eric.dufresne@mat.ethz.ch}

\begin{abstract}

Bicontinuous microstructures are essential to the function of diverse natural and synthetic systems. 
Their synthesis has been based on two
 approaches:  arrested phase separation or self-assembly of block copolymers.
The former is attractive for its chemical simplicity, the latter for its thermodynamic robustness.
Here, we introduce Elastic MicroPhase Separation (EMPS) as an alternative approach to make bicontinuous microstructures.
Conceptually, EMPS  balances the molecular-scale forces that drive demixing with  large-scale elasticity to encode a thermodynamic length scale.
This process features a continuous phase transition, reversible without hysteresis. 
Practically, we trigger EMPS by simply super-saturating an elastomeric matrix with a liquid.
This results in uniform bicontinuous materials with a well-defined microscopic length-scale tuned by the matrix stiffness.
The versatility and robustness of EMPS is further demonstrated by fabricating 
bicontinuous materials with superior mechanical properties and controlled anisotropy and microstructural gradients.

\end{abstract}

\maketitle

Bicontinuous materials feature fascinating interpenetrating structures and  are found across scales in the natural world.
The macroscopic  bicontinuous architecture of marine corals enables an optimal balance between connectivity and surface area for nutrient absorption \cite{CLARKE2011_sponges}. 
In sea urchins, the smoothness of  bicontinuous microstructures leads to  excellent mechanical properties \cite{Yang2022_Urchins}.
At the nanoscale, bicontinuous photonic structures in bird feathers produce beautiful structural colors \cite{dufresne2009self,saranathan2021evolution}.
The unique topology, smooth morphology, and large surface area of these complex materials have inspired the development of synthetic mimicks for applications spanning catalysis \cite{Zielasek2006_catalysis,Li2020_catalysis}, energy storage \cite{Han2023_batteries,Guo2016_batteries}, and optical and mechanical meta-materials \cite{Wohlwend2022_Optical,Shi2021_MetalsMechanics,Biener2006_AuMech}.  
Nonetheless, precise control over bicontinuous microstructures over large volumes presents fundamental challenges.

While bicontinuous structures with domain sizes approaching  tens of microns can be directly written with modern additive manufacturing techniques \cite{Portela2020,Hsieh2019_3Dp},  spontaneous bottom-up approaches are preferred for the efficient production of bicontinuous materials with smaller feature sizes \cite{erlebacher2001evolution,bates201750th,Lu2020_Porous}.  
In polymeric systems, established approaches include arrested phase separation \cite{chan1996polymerization,wienk1996recent}  and the self-assembly of  block copolymers \cite{Chuan2008,xiang2023block}.
While block copolymer assembly offers  unmatched control of bicontinuous structures at macromolecular scales, it cannot produce robust structures at scales much larger than the block copolymer itself.
Conversely,
arrested phase separation uses simpler chemistries and readily produces structures with dimensions well beyond the macromolecular scale \cite{chan1996polymerization,Cates2008_bijels,Guillen_2011_Nonsolvent}. However, these typically compromise the uniformity of the resulting microstructures, as arrest is usually triggered by a diffusion-limited process.
Therefore, the creation of bicontinuous materials with a uniform structure at the microscale remains a major challenge.
To that end, 
matrix elasticity has recently been proposed as an alternative mean to control phase separation \cite{FernandezRico2022}.
While initial experiments have demonstrated precise size tunability at the  scale of several microns \cite{style2018liquid}, the resulting microstructures have been limited to dilute spherical domains that are squeezed out of the matrix over tens of hours \cite{rosowski2020elastic}.

In this paper, we introduce Elastic MicroPhase Separation (EMPS)  to produce bicontinuous microstructures in bulk polymeric materials.
In EMPS, elastic forces imposed by polymer matrices are used to counter molecular-scale forces that drive demixing.
In this way, matrix elasticity stabilizes the microstructure and controls its  final length scale.
Similar to block copolymer assembly, EMPS features a thermodynamically-defined length scale. Like arrested phase separation,
it utilizes simple chemistries and provides access to length scales much larger than its macro-molecular building-blocks. 
Furthermore, we find that EMPS is characterized by an unusual thermodynamic boundary combining features of binodal and spinodal curves.
Finally, we demonstrate the versatility and robustness of EMPS by fabricating bicontinuous materials with enhanced durability, microstructural gradients, and  anisotropy.

\begin{figure*}
\includegraphics[width=\textwidth]{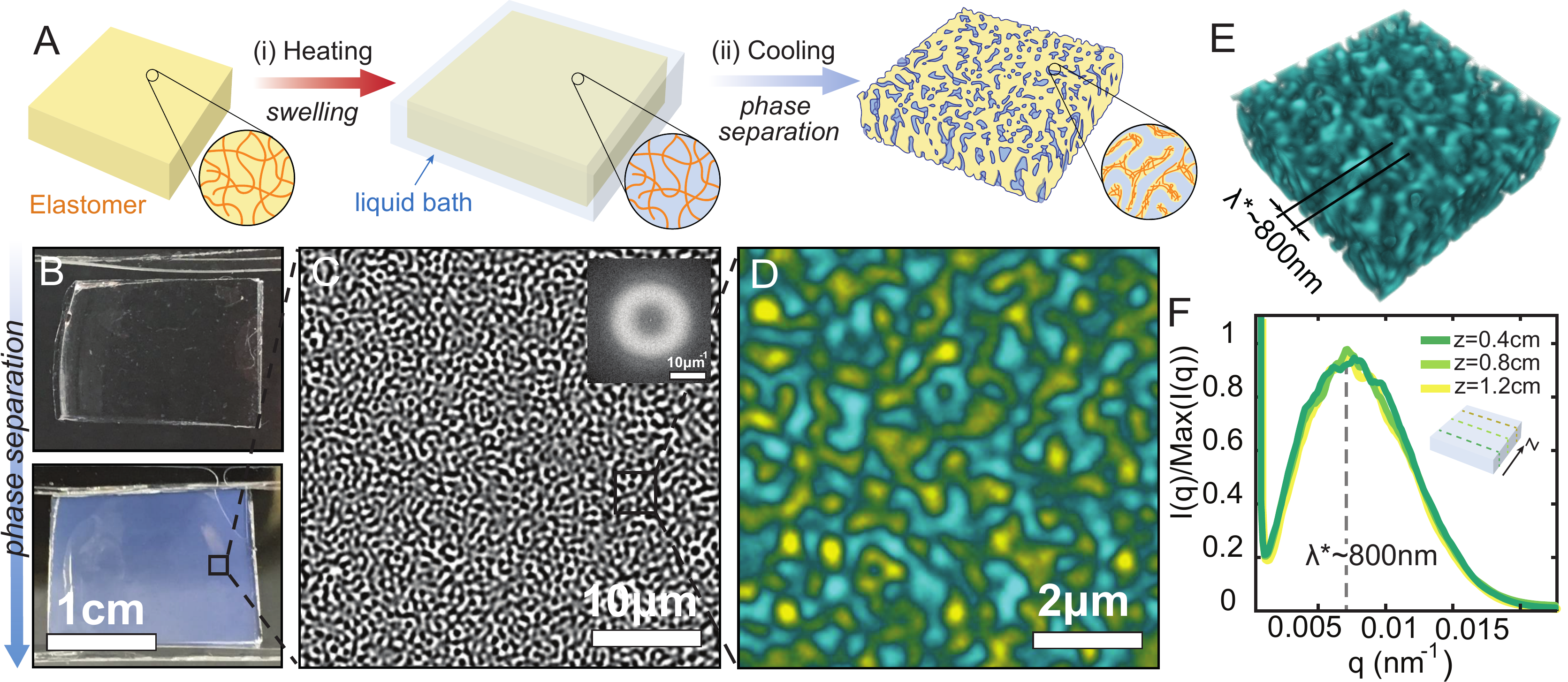}
\caption{Elastic MicroPhase Separation (EMPS) produces bicontinuous microstructures with a well-defined spacing. (A) Schematic diagram of the  process. First, an elastomer is incubated in a bath of liquid at elevated temperatures. After swelling equilibrium is reached, cooling induces phase separation. (B-E) Results for an 800kPa PDMS elastomer swollen with heptafluorobutyl methacrylate (HFBMA) at  T$_{swell}$ = 60$^{\circ}$C (57wt\% of HFBMA) and cooled to room temperature. (B) Macroscopic images of the change of macroscopic colour before (top) and after (bottom) the phase separation process. (C) Bright-field   and (D) confocal microscopy images of microstructure. The inset in C shows the FFT of the shown bright-field image. Yellow domains in (D) depict PDMS-rich domains (Nile red-dyed) and blue domains depict acyrlate-rich domains (BDP-dyed). (E) 3D confocal reconstruction of the bicontinuous structure. (F) Azimuthal average of a 2D FFT analysis of optical microscopy images taken at different cross-sections, along the $z$ axis (see inset). The peak position at $\sim$800~nm does not change across the sample volume (1$\times$2$\times$0.5cm$^3$).}
\end{figure*}

\vspace{-0.6cm} 

\begin{figure*}[]
\includegraphics[width=0.95\textwidth]{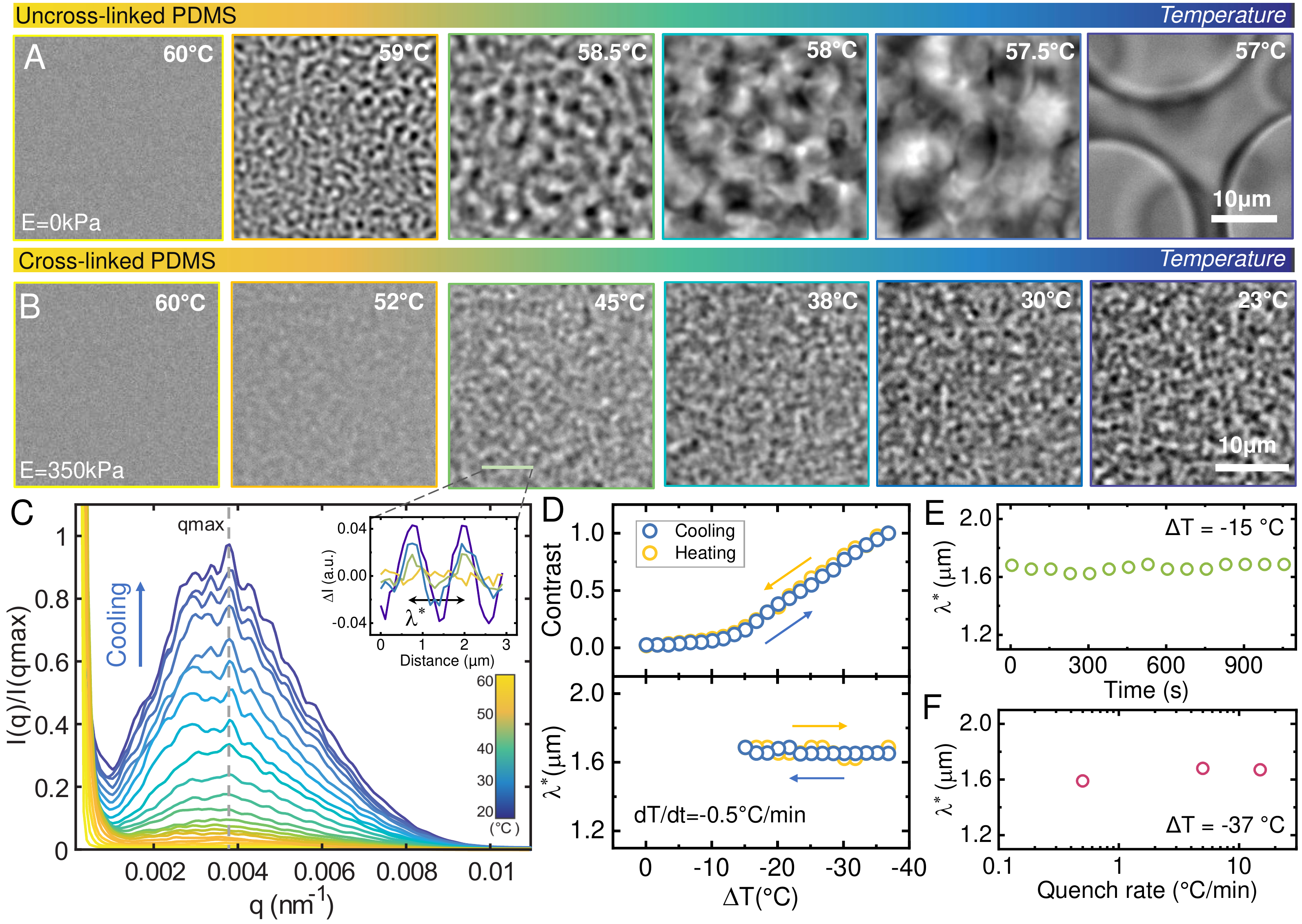}
\caption{EMPS is continuous and produces microstructures with a thermodynamically defined length scale. 
(A,B) Bright-field optical microscopy images of samples cooled from 60$^{\circ}$C to 23$^{\circ}$C at a cooling rate of dT/dt = -0.5$^{\circ}$C/min. (A) Mixtures of HBFMA liquid and uncrosslinked PDMS and (B) mixtures of liquid and crosslinked PDMS (E=350kPa). The final quench depth is $\Delta$T = -37$^{\circ}$C.
(C) Azimuthal average of 2D fast Fourier Transform (FFT) of microscopy images shown in (B). As temperature decreases, a peak in the scattering intensity, $I(q)$, emerges smoothly at a fixed spatial frequency, q$_{max}$. The inset shows line profiles of microscopy images at different temperatures (see color bar), indicating the increase of contrast at a fixed length scale. (D) Contrast ($\mathrm{max}_{q}(I)/\mathrm{max}_{q,T}(I)$)  and characteristic length scale ($\lambda^{*}=2\pi q_{max}^{-1}$) evolution of the bicontinuous microstructure during heating and cooling. (E) Time evolution of $\lambda^{*}$  for a fixed quench depth, $\Delta$T = -15$^{\circ}$C. (F) Quench rate evolution of $\lambda^{*}$ for a fixed $\Delta$T = -37$^{\circ}$C.}
\end{figure*}

\subsection{Super-saturated elastomers form bicontinuous microstructures}

\vspace{-0.4cm}

Our process to fabricate bicontinuous structures is shown schematically in Fig.~1A. 
First, an elastomer is immersed in a bath of liquid at an elevated temperature, T$_{swell}$.
Over the course of a few days, the liquid diffuses into the elastomer, until the system reaches swelling equilibrium. 
Then, the temperature is dropped to reduce the solubility of the liquid in the matrix, inducing Elastic MicroPhase Separation (EMPS). 
An example of an 800kPa polydimethylsiloxane (PDMS) elastomer  swollen with heptafluorobutyl methacrylate (HFBMA) at 60$^\circ$C is shown in Fig.~1B-E.
Macroscopically, the liquid-saturated elastomer is transparent and becomes cloudy blue upon cooling to room temperature (see Fig.~1B).
Bright-field optical microscopy of a thin sample reveals an intricate and regular network structure at the microscale (see Fig.~1C). 
This microstructure features strong structural correlations at a single wavelength, as shown by the ring in the two-dimensional Fourier transform of the image (see inset Fig.~1C). 
Stacks of fluorescence confocal images shown in Fig.1D and E, reveal that the three-dimensional structure of our material is bicontinuous (see full stack in Movie S1).
This highly correlated structure is homogeneous across the centimeter-scale sample, as shown by the Fourier spectra of images acquired at different cross-sections (see Fig.~1F). 
 
The resulting phase-separated structures resemble transient bicontinuous networks formed during  spinodal decomposition \cite{Cahn1958,Cahn1959,Cabral2009}. 
However, in striking contrast to that classical process, our microstructures are stationary for hours without sign of coarsening.
As we show in the following section, the essential ingredient leading to this unusual 
phase separation process 
is the elasticity of the polymer matrix. 
 
\vspace{-0.5cm}
\subsection{Matrix elasticity stabilizes a single length scale}
\vspace{-0.4cm}

\begin{figure*}
\includegraphics[width=\textwidth]{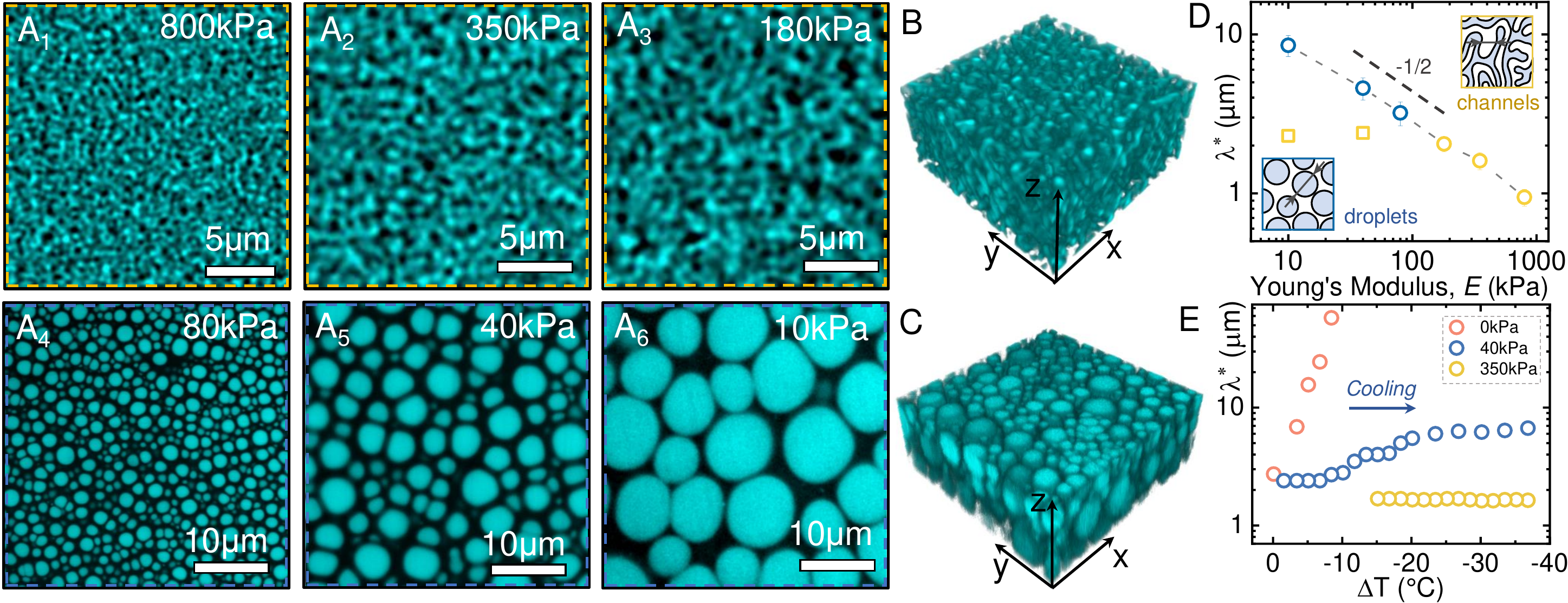} 
\caption{Matrix stiffness tunes microstructural length scale and morphology.
(A) Confocal microscopy images of microphase-separated structures for unswollen PDMS Young's moduli of (A$_{1}$) 800kPa, (A$_{2}$) 350kPa, (A$_{3}$) 180kPa, (A$_{4}$) 80kPa, (A$_{5}$) 40kPa and (A$_{6}$) 10kPa, at 23$^{\circ}$C.  (B,C) 3D reconstructions of microstructures with (B) E=350kPa (14$\times$14$\times$6 $\mu$m$^{3}$) and (C) E=40kPa (40$\times$40 $\times$15 $\mu$m$^{3}$). (D) Characteristic length scale, $\lambda^{*}$, measured by FFT as a function of $E$.  Circles denote results at 23$^{\circ}$C. 
Yellow symbols indicate the channel width of the bicontinuous structures and blue symbols the droplets' diameter. 
Squares represent $\lambda^{*}$ of the bicontinuous structures formed before the droplets.
The error bars represent the standard error of the mean. (E) Evoluation of $\lambda^{*}$  during  cooling  for $E=350$kPa (yellow), $E=40$kPa (blue) and $E=0$kPa (red).}
\label{fig:ps_elasticmedia}
\end{figure*}

To highlight the crucial role of matrix elasticity, we compare phase separation with and without crosslinking of PDMS  (see Fig.~2A and B).
For a fluid matrix with no cross-linking (see Fig.~2A), phase separation follows the classical spinodal decomposition pathway, with the emergence and rapid coarsening of bicontinuous channels, followed by their break-up into discrete droplets \cite{Aarts_2005}.
By contrast, phase separation in cross-linked matrices (see Fig.~2B), results in bicontinuous microstructures that emerge at deeper quenches and do not appear to coarsen over the course of experiment.

Fourier analysis of the image sequence shown in Fig.~2B reveals that the contrast of the structure smoothly increases during cooling, while the characteristic length scale  of the structure remains fixed.
This is shown in Fig.~2C, where the Fourier spectrum at each time-point is characterized by a broad peak centered on a single spatial frequency, $q_{max}$. 
As temperature decreases, the height of this peak (\emph{i.e.} contrast) increases smoothly (see top panel Fig.~2D), while the characteristic length scale  $\lambda^{*}=2\pi q_{max}^{-1}$ remains constant at approximately 1.6~$\mu$m (see bottom panel Fig.~2D). 
Intriguingly, elastic microphase separation is fully reversible without hysteresis when the system is heated up.
This is shown by the yellow data in Fig.~2D, where the contrast and length scale follow the same path for cooling and heating.
We also find that $\lambda^{*}$  is stable over time at a fixed temperature (see Fig.~2E), and that it remains fixed  when the quench rate is varied by a factor of 30 (see Fig.~2F).

 Together, these observations suggest that the contrast and characteristic length scale of the phase-separated microstructure are equilibrium features of the system.
This is reminiscent of microphase separation in block copolymers \cite{bates201750th,Chuan2008}.
However, our microstructures have $\lambda^{*}$ much bigger than the sizes of their constituent macromolecules, which is typical of kinetically arrested phase separation \cite{Gibaud2009,KAHRS2020_nonSolvent,Haase2022}. 
Elastic microphase separation, therefore, combines attractive characteristics of both established routes, and provides a bulk route to produce uniform bicontinous microstructures.

\vspace{-0.6cm}
\subsection{Matrix stiffness controls the length scale and morphology of the microstructure}
\vspace{-0.3cm}

To elucidate the role of elasticity, we vary the matrix stiffness.
Confocal images of the microstructures formed in matrices with Young's moduli, $E$, ranging from 800 to 10kPa, are shown in Fig.~3A.
As $E$ is reduced from 800 to 180kPa, the characteristic length scale of the bicontinuous morphology increases from 0.8 to 2$\mu$m (see Figs.~3A$_{1-3}$ and Fig.~3D).
For softer matrices, the final morphologies are, in fact, dense packings of discrete  droplets, resembling `compressed emulsions' (see Figs. 3A$_{4-6}$) \cite{brujic20033d}. In these droplet structures, the length scale increases from about 3 to 9$\mu$m, when $E$ decreases from 80 to 10kPa (see Fig.~3D). In general, we find that over the full range of stiffnesses, $\lambda^*$ scales smoothly as $1/\sqrt{E}$, as shown by the circles in Fig.~3D (see distributions in Fig.~S2).
Three-dimensional confocal reconstructions of both morphologies are displayed in Figs. 3B and C (see full stacks in Movie S2 and S3).

Independent of the final morphology and stiffness, we find that 
phase separation initiates with the smooth emergence of a bicontinuous structure at a single wavelength (see Fig.~3E and Fig.~S3).  
While this structure is stable in size and connectivity in stiff samples, it coarsens and breaks in softer samples. This is shown in Fig.~3E, where the evolution of $\lambda^{*}$ with cooling is shown for three different elasticities. 
In stiff samples (E=350kPa, Fig.~3E), $\lambda^{*}$ is stable throughout the entire phase separation process. 
In softer samples (E=40kPa, Fig.~3E), $\lambda^{*}$ is initially stable, as a network is formed, but then increases as the structure coarsens and forms droplets at deeper quenches (see images in Fig.~S3). 
The initial characteristic length scale of these networks is added in Fig.~3D as 
 yellow squares. 
 For soft samples, this initial length scale is independent of $E$.
In uncrosslinked matrices (E=0kPa, Fig.~3E), $\lambda^{*}$ rapidly increases as soon as the temperature drops, as expected for classical demixing (see Fig.2A).

To compare EMPS to the early stages of classical spinodal decomposition \cite{Cabral2009}, we constructed a minimal analytical model considering the coupling between demixing and elasticity.
We evaluated the linear stability of a near-critical mixture, following Cahn and Hilliard \cite{Cahn1958}, with an additional linear-elastic term governing stretching of the matrix.
In this model, we find the initial wavelength during the early stage of spinodal demixing is given by $\lambda^{*}= \lambda_0/\sqrt{1 + E/E^*(T)}$. 
Here, $\lambda_0$ is the corresponding wavelength for an uncrosslinked matrix, and E$^*$(T) is a characteristic stiffness depending on the degree of undercooling.
For $E \gg E^*$, we expect the fastest-growing wavelength to scale like $1/\sqrt{E}$, and for $E \ll E^*$, we expect $\lambda^{*}$ to be independent of stiffness.  
This minimal model 
seems to capture  the observed scaling of the initial wavelength $\lambda^{*}$ with $E$ (see yellow data in Fig.~3D).  
However, numerical solutions of this model reveal that linear elasticity is not sufficient to arrest the coarsening of the microstructure over time (see Figs. S11 and S13). 
This suggests that additional factors such as non-linear elasticity and toughness could be playing a role.

\vspace{-0.6cm}
\subsection{Elasticity changes the thermodynamics of phase separation}
\vspace{-0.4cm}

\begin{figure*}
\includegraphics[width=\textwidth]{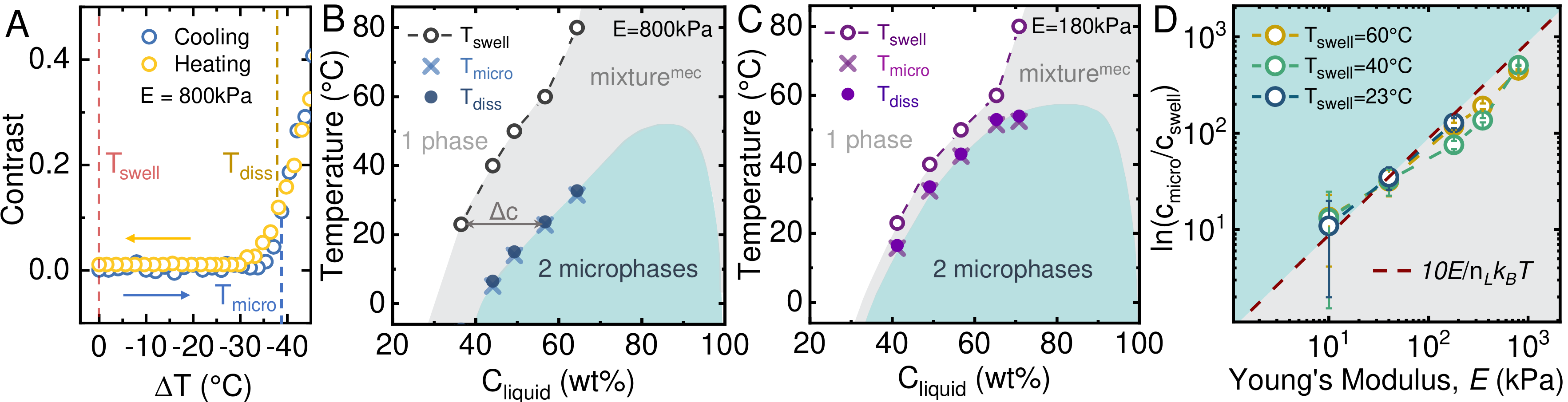}
\caption{Elastic microphase separation is a continuous hysteresis-free  transition.
(A) Contrast ($\mathrm{max}_{q}(I)/\mathrm{max}_{q,T}(I)$)  evolution of an 800kPa sample from 60$^{\circ}$C (T$_{swell}$) to 15$^{\circ}$C at a quench rate $dT/dt=-0.5^{\circ}$C/min. $T_{micro}$ is the temperature at which microphase separation is observed and $T_{diss}$ the temperature at which the microstructure disappears. A contrast threshold of 0.1 is used to determine $T_{micro}$  and $T_{diss}$. (B) Phase diagram of 800kPa samples and (C) 180kPa samples. The boundary at $T_{swell}$ is shown by the empty circles and $T_{micro}$ with the x's. $\Delta$c = c$_{swell}$ - c$_{micro}$ 
 The gray area shows the mechanically-stabilized part of the phase diagram (mixture $^{mec}$) and the blue area where microphase separation is observed (2 microphases). The mixture $^{mec}$ region size increases with $E$. (D) Supersaturation degree of the liquid at the point of microphase separation. Supersaturation increases   approximately as $10E/n_{L}k_{B}T$.
Error bars represent the standard deviation of the mean. 
}
\end{figure*}

While results in the previous section suggest that the microstructure forms by spinodal decomposition, a closer look at the 
phase behavior of EMPS suggests that it is governed by a distinct thermodynamic process.
The key features of our argument are shown in Fig.4~A. 
There, we quantify the structural contrast evolution of a bicontinuous microstructure during a temperature cycle.
Cooling from the binodal curve at $T_{swell}=60^\circ C$,
we observe  no phase separation until $T_{micro} = 21.1^\circ C$ (see blue data Fig.4~A, contrast threshold 0.1), which is $\sim$39$^\circ$C below the swelling temperature. 
Our observations rule out phase separation at $T_{micro}$ as being spinodal decomposition near the critical point, where no gap is expected between the binodal and spinodal curves. 
Spinodal decomposition, however, is still possible away from the critical point.
To test for this, we reheat the sample (see yellow data in Fig.4~A), and find that the microstructure fully dissolves at $T_{diss} = 22^\circ C$, very close to $T_{micro}$. 
This remarkable observation rules out classical off-critical spinodal demixing, where phase-separated domains should persist until the original swelling temperature is reached.

Crucially, we find that $T_{micro} \approx T_{diss}$ is a robust feature across the full range of incubation temperatures (see Fig.~4B) and matrix stiffnesses (see Fig.4BC and Fig.S5). 
Further, $T_{micro}$ is rate independent for sufficiently slow quench rates (see Fig.~S6).
These results suggest that $T_{micro}$ is a proper thermodynamic boundary,
dividing a region where mixtures are mechanically stabilized by the elastic polymer matrix (see highlighted gray zone in Fig.4B-E), from a microphase-separated area (see highlighted cyan zone in Fig.4B-E).
Like a spinodal,  a continuous compositional change is observed when this boundary is crossed. 
Like a binodal, phase-separated domains fully dissolve when it is crossed in the other direction.  
Such continuous hysteresis-free transitions are normally  seen only near critical points. 

\begin{figure*}
\includegraphics[width=0.95\textwidth]{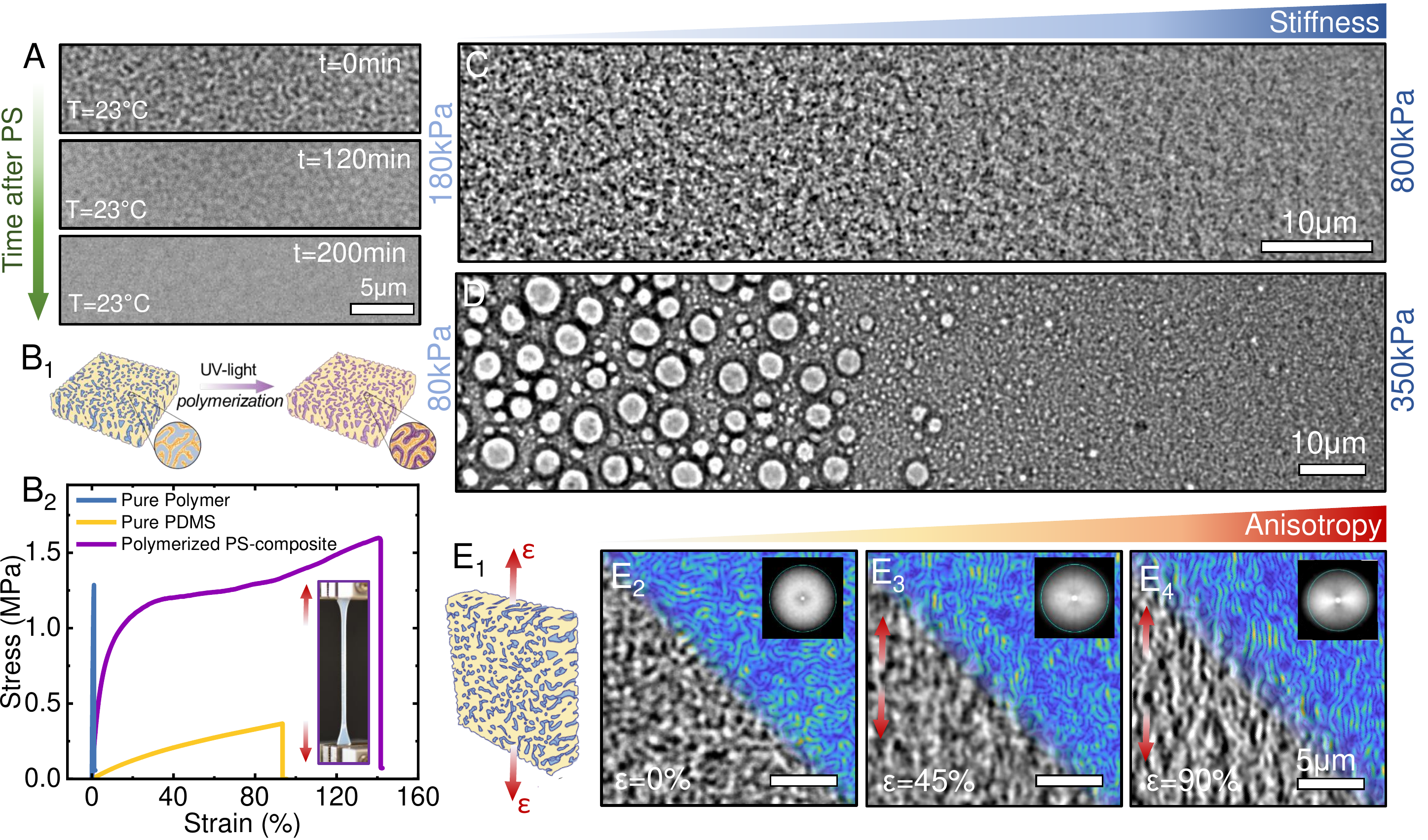}
\caption{EMPS produces versatile microstructures and subsequent polymerization endows durability.
(A) Bright-field microscopy images of the structural evolution a 350kPa sample over time after microphase separation. The sample is kept in contact with the liquid bath.  Without stabilization, contrast slowly disappears near the surface. ($B_{1}$) Polymerization after phase separation stabilizes the microstructure (Fig. S8) and enhances mechanical toughness. ($B_{2}$) Stress-strain curve for pure PDMS (800kPa), pure polymer and polymerized bicontinuous microstructure. The inset shows a photograph  of a tensile test for the polymerized bicontinuous sample. (C,D) Bright-field images of structural gradients created by inducing EMPS in samples with a stiffness gradient. (E) Bright-field images of anisotropic bicontinuous microstructures. Anisotropy is systematically introduced by uniaxially stretching the sample as shown in E$_{1}$. The top-overlays in (E$_{2-4}$) show the bright-field images coloured according to intensity gradient. The insets show the FFT of the respective images.}
\label{Fig3}
\end{figure*}

Matrix stiffness has a strong impact on the  onset of microphase separation.
The gap between the $T_{swell}$ and $T_{micro}$ increases with  matrix stiffness (compare gray zones in Figs.4B-C).  
To interpret this, we  quantify the degree of supersaturation at the point of microphase separation  (see $\Delta$c in Fig.4B) as $\ln(c_{micro}/c_{swell})$ for different elasticities and swelling temperatures.
As shown in Fig.~4D, the supersaturation degree collapses the data acquired at different T$_{swell}$ and scales roughly linearly with $E$ of the unswollen matrix.  
This trend has a simple mechanical interpretation, suggested by \cite{rosowski2020elastic}.
Assuming ideality, the difference in chemical potential between the mixed and marginally microphase separated states is given by $\Delta \mu = k_B T \ln(c_{micro}/c_{swell})$.
This leads to an elevated osmotic pressure  of the solute in the polymer-depleted phase, $\Delta \Pi = n_L \Delta \mu$, which is needed to maintain the host network in a deformed state.
As shown by the red dashed line in Fig.~4D, 
the onset of microscopic phase separation is consistent with $\Delta \Pi \approx 10 E$.
Thus, the competition between elasticity and solubility places a practical upper-bound on the stiffness of samples that can be used, and hence a lower-bound on the length scales accessible by EMPS. 

 \vspace{-0.6cm}
\subsection{EMPS makes functional materials with diverse microstructures}
 \vspace{-0.3cm}

While our bicontinuous materials do not coarsen over time, the elevated osmotic pressure in the oil-rich microphase leads to its slow transport out of the network over several hours (see Fig.~5A) \cite{rosowski2020elastic}. 
However, we can suppress this liquid migration by polymerizing the  methacrylate-functionalized HFBMA after phase separation (see Fig.5B$_{1}$ and Fig.S8).  
We achieve this by adding photo-initiator to the incubation bath and illuminating the sample with UV-light after phase separation (see SM).
The resulting material is permanent (stable for at least 12 months), and thus enables EMPS to be used to fabricate microstructured functional materials.

For instance, EMPS followed by polymerization can significantly increase the toughness and Young's Modulus of elastomers. 
This is shown in Figure 5B$_{2}$, where tensile tests of polymerized samples reveal a remarkably elevated toughness of roughly 2MJ/m$^3$, which is 9$\times$ greater than pure PDMS and 200$\times$ greater than polymerized HFBMA (pHFBMA).
The composite elastomer has a stiffness $\sim$20MPa (11$\times$ greater than pure PDMS) and withstands strains up to $\sim$140\% (see Fig.~5B$_{2}$ and Fig.S9).
This enhanced mechanical performance could arise from a  filler- \cite{frohlich2005_fillerElastomer} or double-network-effect \cite{Gong_DNhydrogels}. Alternatively, the brittle pHFBMA could be plasticized by small quantities of PDMS \cite{Zhang2022}.

Finally, the simplicity and robustness of EMPS enables the fabrication of bicontinuous materials with controlled structural gradients and anisotropy. 
When swelling and phase separation take place in a polymer matrix with a stiffness gradient, the resulting materials can present smooth and continuous changes in both the length scale (see Fig.~5C) and morphology of the microstructure (see Fig.~5D). 
Materials with structural gradients are attractive for filtration, implants or tissue engineering applications, where gradual changes in the length scale are crucial for their functionality \cite{Kumar2022, Gibson2005_cellularSolids}.
Finally, anisotropic bicontinuous structures can be also systematically produced by stretching the material, either before (see Fig.~S10) or after phase separation (see Fig.~5E).
Such anisotropic structures have recently been proposed to have promising orientation-dependent mass transport,
energy absorption, and mechanical properties \cite{Zeng2018_anisotBlock, Engelmayr2008_cardiacAnis, Kumar2022}.

\vspace{-0.6cm}
\subsection{Conclusions}
\vspace{-0.3cm}

Elastic microphase separation is a powerful  approach for fabricating homogeneous bicontinuous materials in bulk. 
The final length scale, morphology, and thermodynamics of this system are found to be intimately related to the mechanics of the elastic matrix.
In fact, we have uncovered the emergence of an elastically-controlled thermodynamic boundary delineating a hysteresis-free continuous phase transition over a wide range of compositions. 
Finally, we have shown the potential of this approach for making functional materials, by fabricating tough elastomers, and bicontinuous materials with controlled anisotropy and microstructural gradients.

Our results have broad implications for soft matter  and materials science. 
For  soft matter, EMPS challenges classical models of phase separation, where  continuous hysteresis-free transitions are only accessible near critical points.
As such, EMPS requires a new theoretical framework, incorporating elasticity to address the stability and length scale of the system.
We anticipate that a deeper understanding of the  interplay  of mechanics and thermodynamics will shed crucial light on emerging problems in  biology \cite{Shin_2006_CondesatesElasticityNucleous,wiegand2020drops,ronceray2022liquid} and mechanical engineering \cite{Portela2020,Kumar2022}.

From a materials perspective, EMPS breaks the current dichotomy between block copolymer self-assembly and arrested phase separation to produce bicontinuous microstructures.
EMPS combines the thermodynamic robustness of block-copolymer assembly with the chemical simplicity of arrested phase separation, enabling precision microstructures with inexpensive components.
While our preliminary results on anisotropic and graded bicontinuous materials already suggest applications in mechanical materials \cite{Kumar2022}, EMPS promises a host of opportunities for developing other functional materials. Further developments of this system could include the reduction of the microstuctural length scale to produce structural color \cite{dufresne2009self}, the optimization of the elastomers' toughness for wearable devices applications \cite{peng2021stiff}, or the selective removal of one of the phases to exploit the exciting filtration, energy storage and catalytic properties of open bicontinuous materials \cite{phillip2010self,Lee2022_batteries,Zielasek2006_catalysis}.

\vspace{-0.4cm}
\subsection{Acknowledgements}
We acknowledge funding from the ETH Zurich Fellowship and the Swiss National Science Foundation NCCR for Bioinspired Materials.  We thank Nan Xue, Kathryn Rosowski, David Zwicker and Dennis Kochmann for useful discussions.

\section*{Methods}
\subsection*{Fabrication of HFBM-PDMS bicontinuous microstructures}
\vspace{-0.3cm}
First, we prepare pure PDMS matrices by mixing PDMS chains (DMS-V31, Gelest) with a crosslinker (HMS-301, Gelest) and a platinum-based catalyst (SIP6831.2, Gelest) (see full recipe in \cite{Style2015}). The stiffness of the matrix depends on the mass ratio between the chains and crosslinker (from 3:1 to 9:1), while keeping the catalyst concentration constant (0.0019\% in volume). Once the different parts are thoroughly mixed together, we pour the mixture into a petri dish, degassed it in vacuum, and finally cure it at 60$^{\circ}$C for approximately 6 days. After curing, the resulting PDMS elastomer is carefully removed from the petri dish and cut into rectangular pieces ($\sim$1cm$\times$2cm$\times$0.5cm). 

Next, PDMS pieces are transferred into a bath of heptafluorobutyl methacrylate (HFBMA, Apollo scientific) ($\sim$1mL HFBMA/0.5g of PDMS), in a 25mL glass bottle. This bath is then typically incubated at $T_{swell } = 60^{\circ}$C in a pre-heated oven during 2.5 days. After the elastomer is saturated with the liquid, the glass bottle is brought to room temperature, at which point phase separation occurs spontaneously. The resulting phase-separated bicontiunous samples are then prepared for characterization tests.

Note that experiments with un-crosslinked PDMS are performed by preparing mixtures of PDMS chains and HFBMA liquid in glass bottles, and following the same temperature steps described above.
\vspace{-0.3cm}

\subsection*{Structural characterization}
\vspace{-0.3cm}
Bright-field optical and confocal fluorescence microscopy images of phase-separated microstructures are obtained using a Nikon-Ti Eclipse inverted optical microscope, equipped with a Confocal Spinning disk Scanner Unit. We typically use 60$\times$ and 100$\times$ oil objectives with NA=1.2 and 1.45, respectively. Optimal imaging is achieved when preparing thin PDMS samples ($\sim$200-500$\mu$m thick), by either curing the PDMS in between two coverslips (1.5, Menzel Glaser), or by using a razor blade to cut thin slides from the bulk samples. Note that the incubation period for the samples between coverslips is significanly longer ($\sim$6 days). For confocal microsocopy experiments,  free BDP is added to HFBMA and Nile-Red is added to PDMS prior to incubation. 

To characterize the structural evolution of the samples with temperature we use thin PDMS samples ($\sim$200-500$\mu$m thick) cured on a 35-mm diameter glass-bottomed dish (MatTek). We add 1mL of HFBMA, and seal the petri dish with Teflon tape. The sample is next incubated at 60$^{\circ}$C-oven for 2.5 days. The sample is next transferred into a pre-heated heating stage at 60$^{\circ}$C (InsTec instruments), which is coupled to an optical microscope. For these experiments we use an 60$\times$ air objective, to avoid temperature variations in the sample.

\vspace{-0.5cm}
\subsection*{Stiffness-gradient and anisotropic samples}
\vspace{-0.3cm}
To prepare PDMS samples with stiffness gradients, we first cure a stiff PDMS mixture (\textit{e.g.}~3:1 chains:crosslinker) on the right-side of a coverslip, and let it cure at 40$^{\circ}$C for 2~hours. We limit it to the right-side by only pouring a small amount of mixture, and using a very small tilt during the curing.
Next, we cure a softer PDMS mixture (\textit{e.g.}~5:1 chains:crosslinker) on the left-side of the coverslip, which slowly gets in contact with the pre-cured stiff side. 
The gradient is let to cure for two days at 60$^\circ$C.

Anisotropic bicontinuous structures can be prepared by either pre-stretching a PDMS sample before the incubation and phase separation process, or after the system has phase separated. In both cases, strains are externally imposed by clamping 8cm$\times$2cm$\times$0.5cm pieces of the elastomer, in a home-made device for stretching the sample that can be coupled to an optical microscope (see Fig.~S10). The stretch ($\epsilon$) is calculated as $\Delta$l/l$_{0}$, where l$_{0}$ is the original unstretched length. Stretch can be applied before incubation or after phase separation.
\vspace{-0.3cm}
\subsection*{Polymerization of bicontinuous microstructures and mechanical tests}
\vspace{-0.3cm}
Polymerization of the phase separated samples is performed in degassed glass vials containing a piece of PDMS ($\sim$0.4g), 500$\mu$L of HFBMA and 2wt\% of photoinitiator (2-Hydroxy-2-methylpropiophenone, Apollo scientific). Once the incubation has proceed at 60$^{\circ}$C for 2.5 days, the samples are cooled down to room temperature to induce EMPS. Next, we remove the excess of monomer in the container with a degassed syringe and expose the sample to UV-light for 2 hours. We use a 365nm UV-lamp (12 Watts and 0.5cm from the lamp).

To measure the mechanical properties of the polymerized materials, we prepare 5$\times$2$\times$0.5cm$^3$ dog-bone PDMS samples, and proceed with the swelling, phase separation and polymerization procedure described above. Once dog-bones are polymerized, we perform uniaxial tests using a tensile testing machine (Stable Micro Systems), where we record the engineering strain-stress curves until failure. All mechanical tests were performed in air, at room temperature, and at elongation speeds of 0.05mm per second. The sample toughness (J/m$^{3}$) is calculated as the area under the strain-stress curves.

\vspace{-0.3cm}

\subsection*{Thermo-mechanical model based on Cahn-Hilliard theory}
\vspace{-0.3cm}

Assuming  spinodal decomposition near a critical point, we introduce a minimal model with two basic ingredients, a Landau-Ginzburg free energy \cite{Landauginzburg} and an energy term related to the mechanical properties of the matrix.
We define $\phi(x)$ as an order parameter describing how far we are from the critical point ($\phi=0$).
The matrix has a stiffness $E_{sw}(\phi)=E_0(1-m\phi)$, where $E_0$ is the Young's modulus at the critical point, and $m$ captures the compositional variation of the stiffness. For simplicity, we assume $m$ to be independent of $\phi$ and $E_0$.
Then,  free energy density including elasticity is represented as:
\begin{equation}
f(\phi,\epsilon)=-\frac{1}{2} \alpha(T) \phi^2 + \beta \phi^4 + \frac{1}{2}\kappa \phi'^2+\frac{1}{2} E_{sw}(\phi) \epsilon^2 - E_0 \epsilon_0\epsilon.
\end{equation}
The first two terms on the right side represent the free energy of mixing of the matrix and the liquid, while the third represents the interfacial energy between the two materials.
$\alpha(T)$, $\beta$ and $\kappa$ are the usual coefficients for a Landau-Ginzburg type free energy, and we take these to be independent of the network elasticity. 
The last two terms represent the mechanical strain energy of the mixture, which includes the stored elastic strain energy and a contribution related to the swelling pressure that makes the matrix equilibrate at a non-zero strain, $\epsilon_0$.

 We assume the phase separation processes maintains local mechanical equilibrium. This equilibrium condition is given by $\partial f/\partial \epsilon=0$, so $E_{sw}(\phi)\epsilon=E_0\epsilon_0$.
Thermodynamic equilibrium is much slower to achieve, as it involves slow transport of fluid through the sample.
This transport occurs along gradients in the exchange chemical potential \cite{konig2021two}:
\begin{align}
\mu&=\left.\frac{\partial f}{\partial \phi}\right|_{\epsilon,\phi'} -\frac{d}{dx}\left(\left.\frac{\partial f}{\partial \phi'}\right|_{\epsilon,\phi}\right)\\
       &=-\alpha(T)\phi + 4 \beta \phi^3+\frac{E_0^2\epsilon_0^2 E_{sw}'(\phi)}{2E_{sw}^2(\phi)} - \kappa \phi''.\nonumber
\end{align}


The flux of monomer is given by $J=-Md\mu/dx$, where $M$ is the mobility, which we assume to be constant. Conservation of mass is described by the transport equation:
\begin{equation}
\frac{\partial \phi}{\partial t}=-\frac{dJ}{dx}=M\frac{d^2\mu}{dx^2}.
\end{equation}

To find the wavelength that emerges at the onset of phase separation, we perform a linear-stability analysis of this equation \cite{Cahn1958}.
We let $\phi=0+\delta \phi$, etc.
Then, at leading order, we find the linearized transport equation:
\begin{equation}
\begin{split}
\frac{1}{M}\frac{\partial \delta \phi}{\partial t} &=\delta \phi'' \left[-\alpha(T)-\epsilon_0^2E_0 m^2\right]-\kappa \delta \phi'''' \\ &\equiv h_0 \delta\phi''-\kappa \delta\phi''''.
\end{split}
\end{equation}
$h_0$ is defined as the term in the square brackets for convenience. We seek solutions of the form $\delta \phi = Ae^{-ikx+\omega t}$, where $\omega$ is the growth rate of a mode with wavelength $2\pi/k$. Inserting this into the transport equation, we find
\begin{equation}
\frac{\omega}{M}=-k^2 h_0-k^4\kappa.
\end{equation}
When $\omega$ is positive for some range of wavelengths, the system is unstable, and spinodal decomposition will occur.
This occurs when $h_0<0$ (\emph{n.b.} $\kappa$ is always positive).
Under these conditions, we see that $\omega$ will take a maximum value when $d\omega/dk=0$. The corresponding value of $k$ gives us the wavelength which will dominate the early stages of spinodal decomposition:
\begin{equation}
\lambda^*=2 \pi\sqrt{\frac{2\kappa}{-h_0}} =2 \pi\sqrt{2\kappa/\left[\alpha(T)+\epsilon_0^2E_0 m^2\right]}.
\end{equation}

Assuming $m$ and $\epsilon_0$ to be independent of experimental conditions, and $E_0$ to be proportional to unswollen Young's modulus of the matrix, $E$, we can write the expression above as
\begin{equation}
\lambda^*\approx 2 \pi\sqrt{\frac{2\kappa}{\alpha(T)+a E}}=\frac{\lambda_0}{\sqrt{1+E/E^*(T)}},
\end{equation}
where $a$ is a constant, $\lambda_0=2\pi\sqrt{2\kappa/\alpha(T)}$, and $E^*(T)=\alpha(T)/a$.

\subsection{References}

\begin{mcitethebibliography}{49}
\providecommand*\natexlab[1]{#1}
\providecommand*\mciteSetBstSublistMode[1]{}
\providecommand*\mciteSetBstMaxWidthForm[2]{}
\providecommand*\mciteBstWouldAddEndPuncttrue
  {\def\EndOfBibitem{\unskip.}}
\providecommand*\mciteBstWouldAddEndPunctfalse
  {\let\EndOfBibitem\relax}
\providecommand*\mciteSetBstMidEndSepPunct[3]{}
\providecommand*\mciteSetBstSublistLabelBeginEnd[3]{}
\providecommand*\EndOfBibitem{}
\mciteSetBstSublistMode{f}
\mciteSetBstMaxWidthForm{subitem}{(\alph{mcitesubitemcount})}
\mciteSetBstSublistLabelBeginEnd
  {\mcitemaxwidthsubitemform\space}
  {\relax}
  {\relax}

\bibitem[Clarke \latin{et~al.}(2011)Clarke, Walsh, Maggs, and
  Buchanan]{CLARKE2011_sponges}
Clarke,~S.; Walsh,~P.; Maggs,~C.; Buchanan,~F. Designs from the deep: Marine
  organisms for bone tissue engineering. \emph{Biotechnology Advances}
  \textbf{2011}, \emph{29}, 610--617\relax
\mciteBstWouldAddEndPuncttrue
\mciteSetBstMidEndSepPunct{\mcitedefaultmidpunct}
{\mcitedefaultendpunct}{\mcitedefaultseppunct}\relax
\EndOfBibitem
\bibitem[Yang \latin{et~al.}(2022)Yang, Jia, Wu, Chen, Deng, Chen, Zhu, and
  Li]{Yang2022_Urchins}
Yang,~T.; Jia,~Z.; Wu,~Z.; Chen,~H.; Deng,~Z.; Chen,~L.; Zhu,~Y.; Li,~L. High
  strength and damage-tolerance in echinoderm stereom as a natural bicontinuous
  ceramic cellular solid. \emph{Nat Commun} \textbf{2022}, \emph{13},
  610--617\relax
\mciteBstWouldAddEndPuncttrue
\mciteSetBstMidEndSepPunct{\mcitedefaultmidpunct}
{\mcitedefaultendpunct}{\mcitedefaultseppunct}\relax
\EndOfBibitem
\bibitem[Dufresne \latin{et~al.}(2009)Dufresne, Noh, Saranathan, Mochrie, Cao,
  and Prum]{dufresne2009self}
Dufresne,~E.~R.; Noh,~H.; Saranathan,~V.; Mochrie,~S.~G.; Cao,~H.; Prum,~R.~O.
  Self-assembly of amorphous biophotonic nanostructures by phase separation.
  \emph{Soft Matter} \textbf{2009}, \emph{5}, 1792--1795\relax
\mciteBstWouldAddEndPuncttrue
\mciteSetBstMidEndSepPunct{\mcitedefaultmidpunct}
{\mcitedefaultendpunct}{\mcitedefaultseppunct}\relax
\EndOfBibitem
\bibitem[Saranathan \latin{et~al.}(2021)Saranathan, Narayanan, Sandy, Dufresne,
  and Prum]{saranathan2021evolution}
Saranathan,~V.; Narayanan,~S.; Sandy,~A.; Dufresne,~E.~R.; Prum,~R.~O.
  Evolution of single gyroid photonic crystals in bird feathers.
  \emph{Proceedings of the National Academy of Sciences} \textbf{2021},
  \emph{118}\relax
\mciteBstWouldAddEndPuncttrue
\mciteSetBstMidEndSepPunct{\mcitedefaultmidpunct}
{\mcitedefaultendpunct}{\mcitedefaultseppunct}\relax
\EndOfBibitem
\bibitem[Zielasek \latin{et~al.}(2006)Zielasek, Jürgens, Schulz, Biener,
  Biener, Hamza, and Bäumer]{Zielasek2006_catalysis}
Zielasek,~V.; Jürgens,~B.; Schulz,~C.; Biener,~J.; Biener,~M.~M.;
  Hamza,~A.~V.; Bäumer,~M. Gold Catalysts: Nanoporous Gold Foams.
  \emph{Angewandte Chemie International Edition} \textbf{2006}, \emph{45},
  8241--8244\relax
\mciteBstWouldAddEndPuncttrue
\mciteSetBstMidEndSepPunct{\mcitedefaultmidpunct}
{\mcitedefaultendpunct}{\mcitedefaultseppunct}\relax
\EndOfBibitem
\bibitem[Li \latin{et~al.}(2020)Li, Chen, Li, Liu, Bi, Zhang, Zhou, and
  Mai]{Li2020_catalysis}
Li,~Q.; Chen,~C.; Li,~C.; Liu,~R.; Bi,~S.; Zhang,~P.; Zhou,~Y.; Mai,~Y. Ordered
  Bicontinuous Mesoporous Polymeric Semiconductor Photocatalyst. \emph{ACS
  Nano} \textbf{2020}, \emph{14}, 13652--13662\relax
\mciteBstWouldAddEndPuncttrue
\mciteSetBstMidEndSepPunct{\mcitedefaultmidpunct}
{\mcitedefaultendpunct}{\mcitedefaultseppunct}\relax
\EndOfBibitem
\bibitem[Han \latin{et~al.}(2023)Han, Lee, Lee, Lee, Kwon, Min, Lee, Lee, Lee,
  and Kim]{Han2023_batteries}
Han,~J.; Lee,~M.~J.; Lee,~K.; Lee,~Y.~J.; Kwon,~S.~H.; Min,~J.~H.; Lee,~E.;
  Lee,~W.; Lee,~S.~W.; Kim,~B.~J. Role of Bicontinuous Structure in Elastomeric
  Electrolytes for High-Energy Solid-State Lithium-Metal Batteries.
  \emph{Advanced Materials} \textbf{2023}, \emph{35}, 2205194\relax
\mciteBstWouldAddEndPuncttrue
\mciteSetBstMidEndSepPunct{\mcitedefaultmidpunct}
{\mcitedefaultendpunct}{\mcitedefaultseppunct}\relax
\EndOfBibitem
\bibitem[Guo \latin{et~al.}(2016)Guo, Han, Liu, Chen, Ito, Jian, Jin, Hirata,
  Li, Fujita, Asao, Zhou, and Chen]{Guo2016_batteries}
Guo,~X.; Han,~J.; Liu,~P.; Chen,~L.; Ito,~Y.; Jian,~Z.; Jin,~T.; Hirata,~A.;
  Li,~F.; Fujita,~T.; Asao,~N.; Zhou,~H.; Chen,~M. Hierarchical nanoporosity
  enhanced reversible capacity of bicontinuous nanoporous metal based Li-O2
  battery. \emph{Scientific Reports} \textbf{2016}, \emph{6}, 33466\relax
\mciteBstWouldAddEndPuncttrue
\mciteSetBstMidEndSepPunct{\mcitedefaultmidpunct}
{\mcitedefaultendpunct}{\mcitedefaultseppunct}\relax
\EndOfBibitem
\bibitem[Wohlwend \latin{et~al.}(2022)Wohlwend, Sologubenko, Döbeli, Galinski,
  and Spolenak]{Wohlwend2022_Optical}
Wohlwend,~J.; Sologubenko,~A.~S.; Döbeli,~M.; Galinski,~H.; Spolenak,~R.
  Chemical Engineering of Cu–Sn Disordered Network Metamaterials. \emph{Nano
  Letters} \textbf{2022}, \emph{22}, 853--859\relax
\mciteBstWouldAddEndPuncttrue
\mciteSetBstMidEndSepPunct{\mcitedefaultmidpunct}
{\mcitedefaultendpunct}{\mcitedefaultseppunct}\relax
\EndOfBibitem
\bibitem[Shi \latin{et~al.}(2021)Shi, Li, Ngo-Dinh, Markmann, and
  Weissmüller]{Shi2021_MetalsMechanics}
Shi,~S.; Li,~Y.; Ngo-Dinh,~B.-N.; Markmann,~J.; Weissmüller,~J. Scaling
  behavior of stiffness and strength of hierarchical network nanomaterials.
  \emph{Science} \textbf{2021}, \emph{371}, 1026--1033\relax
\mciteBstWouldAddEndPuncttrue
\mciteSetBstMidEndSepPunct{\mcitedefaultmidpunct}
{\mcitedefaultendpunct}{\mcitedefaultseppunct}\relax
\EndOfBibitem
\bibitem[Biener \latin{et~al.}(2006)Biener, Hodge, Hayes, Volkert, Zepeda-Ruiz,
  Hamza, and Abraham]{Biener2006_AuMech}
Biener,~J.; Hodge,~A.~M.; Hayes,~J.~R.; Volkert,~C.~A.; Zepeda-Ruiz,~L.~A.;
  Hamza,~A.~V.; Abraham,~F.~F. Size Effects on the Mechanical Behavior of
  Nanoporous Au. \emph{Nano Letters} \textbf{2006}, \emph{6}, 2379--2382\relax
\mciteBstWouldAddEndPuncttrue
\mciteSetBstMidEndSepPunct{\mcitedefaultmidpunct}
{\mcitedefaultendpunct}{\mcitedefaultseppunct}\relax
\EndOfBibitem
\bibitem[Portela \latin{et~al.}(2020)Portela, Vidyasagar, Krödel, Weissenbach,
  Yee, Greer, and Kochmann]{Portela2020}
Portela,~C.~M.; Vidyasagar,~A.; Krödel,~S.; Weissenbach,~T.; Yee,~D.~W.;
  Greer,~J.~R.; Kochmann,~D.~M. Extreme mechanical resilience of self-assembled
  nanolabyrinthine materials. \emph{Proceedings of the National Academy of
  Sciences} \textbf{2020}, \emph{117}, 5686--5693\relax
\mciteBstWouldAddEndPuncttrue
\mciteSetBstMidEndSepPunct{\mcitedefaultmidpunct}
{\mcitedefaultendpunct}{\mcitedefaultseppunct}\relax
\EndOfBibitem
\bibitem[Hsieh \latin{et~al.}(2019)Hsieh, Endo, Zhang, Bauer, and
  Valdevit]{Hsieh2019_3Dp}
Hsieh,~M.-T.; Endo,~B.; Zhang,~Y.; Bauer,~J.; Valdevit,~L. The mechanical
  response of cellular materials with spinodal topologies. \emph{Journal of the
  Mechanics and Physics of Solids} \textbf{2019}, \emph{125}, 401--419\relax
\mciteBstWouldAddEndPuncttrue
\mciteSetBstMidEndSepPunct{\mcitedefaultmidpunct}
{\mcitedefaultendpunct}{\mcitedefaultseppunct}\relax
\EndOfBibitem
\bibitem[Erlebacher \latin{et~al.}(2001)Erlebacher, Aziz, Karma, Dimitrov, and
  Sieradzki]{erlebacher2001evolution}
Erlebacher,~J.; Aziz,~M.~J.; Karma,~A.; Dimitrov,~N.; Sieradzki,~K. Evolution
  of nanoporosity in dealloying. \emph{Nature} \textbf{2001}, \emph{410},
  450--453\relax
\mciteBstWouldAddEndPuncttrue
\mciteSetBstMidEndSepPunct{\mcitedefaultmidpunct}
{\mcitedefaultendpunct}{\mcitedefaultseppunct}\relax
\EndOfBibitem
\bibitem[Bates and Bates(2017)Bates, and Bates]{bates201750th}
Bates,~C.~M.; Bates,~F.~S. 50th Anniversary Perspective: Block Polymers-- Pure
  Potential. \emph{Macromolecules} \textbf{2017}, \emph{50}, 3--22\relax
\mciteBstWouldAddEndPuncttrue
\mciteSetBstMidEndSepPunct{\mcitedefaultmidpunct}
{\mcitedefaultendpunct}{\mcitedefaultseppunct}\relax
\EndOfBibitem
\bibitem[Xuanming~Lu(2020)]{Lu2020_Porous}
Xuanming~Lu,~K. K. . K.~N.,~George~Hasegawa Hierarchically porous monoliths
  prepared via sol–gel process accompanied by spinodal decomposition. \emph{J
  Sol-Gel Sci Technol} \textbf{2020}, \emph{95}, 530–550\relax
\mciteBstWouldAddEndPuncttrue
\mciteSetBstMidEndSepPunct{\mcitedefaultmidpunct}
{\mcitedefaultendpunct}{\mcitedefaultseppunct}\relax
\EndOfBibitem
\bibitem[Chan and Rey(1996)Chan, and Rey]{chan1996polymerization}
Chan,~P.~K.; Rey,~A.~D. Polymerization-induced phase separation. 1. Droplet
  size selection mechanism. \emph{Macromolecules} \textbf{1996}, \emph{29},
  8934--8941\relax
\mciteBstWouldAddEndPuncttrue
\mciteSetBstMidEndSepPunct{\mcitedefaultmidpunct}
{\mcitedefaultendpunct}{\mcitedefaultseppunct}\relax
\EndOfBibitem
\bibitem[Wienk \latin{et~al.}(1996)Wienk, Boom, Beerlage, Bulte, Smolders, and
  Strathmann]{wienk1996recent}
Wienk,~I.; Boom,~R.; Beerlage,~M.; Bulte,~A.; Smolders,~C.; Strathmann,~H.
  Recent advances in the formation of phase inversion membranes made from
  amorphous or semi-crystalline polymers. \emph{Journal of membrane science}
  \textbf{1996}, \emph{113}, 361--371\relax
\mciteBstWouldAddEndPuncttrue
\mciteSetBstMidEndSepPunct{\mcitedefaultmidpunct}
{\mcitedefaultendpunct}{\mcitedefaultseppunct}\relax
\EndOfBibitem
\bibitem[Tang \latin{et~al.}(2008)Tang, Lennon, Fredrickson, Kramer, and
  Hawker]{Chuan2008}
Tang,~C.; Lennon,~E.~M.; Fredrickson,~G.~H.; Kramer,~E.~J.; Hawker,~C.~J.
  Evolution of block copolymer lithography to highly ordered square arrays.
  \emph{Science} \textbf{2008}, \emph{322}, 429--432\relax
\mciteBstWouldAddEndPuncttrue
\mciteSetBstMidEndSepPunct{\mcitedefaultmidpunct}
{\mcitedefaultendpunct}{\mcitedefaultseppunct}\relax
\EndOfBibitem
\bibitem[Xiang \latin{et~al.}(2023)Xiang, Li, Li, Yang, Xu, and
  Mai]{xiang2023block}
Xiang,~L.; Li,~Q.; Li,~C.; Yang,~Q.; Xu,~F.; Mai,~Y. Block copolymer
  self-assembly directed synthesis of porous materials with ordered
  bicontinuous structures and their potential applications. \emph{Advanced
  Materials} \textbf{2023}, \emph{35}, 2207684\relax
\mciteBstWouldAddEndPuncttrue
\mciteSetBstMidEndSepPunct{\mcitedefaultmidpunct}
{\mcitedefaultendpunct}{\mcitedefaultseppunct}\relax
\EndOfBibitem
\bibitem[Cates and Clegg(2008)Cates, and Clegg]{Cates2008_bijels}
Cates,~M.~E.; Clegg,~P.~S. Bijels: a new class of soft materials. \emph{Soft
  Matter} \textbf{2008}, \emph{4}, 2132--2138\relax
\mciteBstWouldAddEndPuncttrue
\mciteSetBstMidEndSepPunct{\mcitedefaultmidpunct}
{\mcitedefaultendpunct}{\mcitedefaultseppunct}\relax
\EndOfBibitem
\bibitem[Guillen \latin{et~al.}(2011)Guillen, Pan, Li, and
  Hoek]{Guillen_2011_Nonsolvent}
Guillen,~G.~R.; Pan,~Y.; Li,~M.; Hoek,~E. M.~V. Preparation and
  Characterization of Membranes Formed by Nonsolvent Induced Phase Separation:
  A Review. \emph{Cell} \textbf{2011}, \emph{175}, 3798\relax
\mciteBstWouldAddEndPuncttrue
\mciteSetBstMidEndSepPunct{\mcitedefaultmidpunct}
{\mcitedefaultendpunct}{\mcitedefaultseppunct}\relax
\EndOfBibitem
\bibitem[Fernández-Rico \latin{et~al.}(2022)Fernández-Rico, Sai, Sicher,
  Style, and Dufresne]{FernandezRico2022}
Fernández-Rico,~C.; Sai,~T.; Sicher,~A.; Style,~R.~W.; Dufresne,~E.~R. Putting
  the Squeeze on Phase Separation. \emph{JACS Au} \textbf{2022}, \emph{2},
  66--73\relax
\mciteBstWouldAddEndPuncttrue
\mciteSetBstMidEndSepPunct{\mcitedefaultmidpunct}
{\mcitedefaultendpunct}{\mcitedefaultseppunct}\relax
\EndOfBibitem
\bibitem[Style \latin{et~al.}(2018)Style, Sai, Fanelli, Ijavi,
  Smith-Mannschott, Xu, Wilen, and Dufresne]{style2018liquid}
Style,~R.~W.; Sai,~T.; Fanelli,~N.; Ijavi,~M.; Smith-Mannschott,~K.; Xu,~Q.;
  Wilen,~L.~A.; Dufresne,~E.~R. Liquid-liquid phase separation in an elastic
  network. \emph{Physical Review X} \textbf{2018}, \emph{8}, 011028\relax
\mciteBstWouldAddEndPuncttrue
\mciteSetBstMidEndSepPunct{\mcitedefaultmidpunct}
{\mcitedefaultendpunct}{\mcitedefaultseppunct}\relax
\EndOfBibitem
\bibitem[Rosowski \latin{et~al.}(2020)Rosowski, Sai, Vidal-Henriquez, Zwicker,
  Style, and Dufresne]{rosowski2020elastic}
Rosowski,~K.~A.; Sai,~T.; Vidal-Henriquez,~E.; Zwicker,~D.; Style,~R.~W.;
  Dufresne,~E.~R. Elastic ripening and inhibition of liquid–liquid phase
  separation. \emph{Nature Physics} \textbf{2020}, \emph{16}, 422--425\relax
\mciteBstWouldAddEndPuncttrue
\mciteSetBstMidEndSepPunct{\mcitedefaultmidpunct}
{\mcitedefaultendpunct}{\mcitedefaultseppunct}\relax
\EndOfBibitem
\bibitem[Cahn and Hilliard(1958)Cahn, and Hilliard]{Cahn1958}
Cahn,~J.~W.; Hilliard,~J.~E. Free Energy of a Nonuniform System. I. Interfacial
  Free Energy. \emph{The Journal of Chemical Physics} \textbf{1958}, \emph{28},
  258--267\relax
\mciteBstWouldAddEndPuncttrue
\mciteSetBstMidEndSepPunct{\mcitedefaultmidpunct}
{\mcitedefaultendpunct}{\mcitedefaultseppunct}\relax
\EndOfBibitem
\bibitem[Cahn(1959)]{Cahn1959}
Cahn,~J.~W. Free Energy of a Nonuniform System. II. Thermodynamic Basis.
  \emph{The Journal of Chemical Physics} \textbf{1959}, \emph{30},
  1121--1124\relax
\mciteBstWouldAddEndPuncttrue
\mciteSetBstMidEndSepPunct{\mcitedefaultmidpunct}
{\mcitedefaultendpunct}{\mcitedefaultseppunct}\relax
\EndOfBibitem
\bibitem[Cabral and Higgins(2018)Cabral, and Higgins]{Cabral2009}
Cabral,~J.~T.; Higgins,~J.~S. Spinodal nanostructures in polymer blends: On the
  validity of the Cahn-Hilliard length scale prediction. \emph{Progress in
  Polymer Science} \textbf{2018}, \emph{81}, 1--21\relax
\mciteBstWouldAddEndPuncttrue
\mciteSetBstMidEndSepPunct{\mcitedefaultmidpunct}
{\mcitedefaultendpunct}{\mcitedefaultseppunct}\relax
\EndOfBibitem
\bibitem[Aarts \latin{et~al.}(2005)Aarts, Dullens, and
  Lekkerkerker]{Aarts_2005}
Aarts,~D. G. A.~L.; Dullens,~R. P.~A.; Lekkerkerker,~H. N.~W. Interfacial
  dynamics in demixing systems with ultralow interfacial tension. \emph{New
  Journal of Physics} \textbf{2005}, \emph{7}, 40\relax
\mciteBstWouldAddEndPuncttrue
\mciteSetBstMidEndSepPunct{\mcitedefaultmidpunct}
{\mcitedefaultendpunct}{\mcitedefaultseppunct}\relax
\EndOfBibitem
\bibitem[Gibaud and Schurtenberger(2009)Gibaud, and Schurtenberger]{Gibaud2009}
Gibaud,~T.; Schurtenberger,~P. A closer look at arrested spinodal decomposition
  in protein solutions. \emph{Journal of Physics: Condensed Matter}
  \textbf{2009}, \emph{21}, 322201\relax
\mciteBstWouldAddEndPuncttrue
\mciteSetBstMidEndSepPunct{\mcitedefaultmidpunct}
{\mcitedefaultendpunct}{\mcitedefaultseppunct}\relax
\EndOfBibitem
\bibitem[Kahrs and Schwellenbach(2020)Kahrs, and
  Schwellenbach]{KAHRS2020_nonSolvent}
Kahrs,~C.; Schwellenbach,~J. Membrane formation via non-solvent induced phase
  separation using sustainable solvents: A comparative study. \emph{Polymer}
  \textbf{2020}, \emph{186}, 122071\relax
\mciteBstWouldAddEndPuncttrue
\mciteSetBstMidEndSepPunct{\mcitedefaultmidpunct}
{\mcitedefaultendpunct}{\mcitedefaultseppunct}\relax
\EndOfBibitem
\bibitem[Khan \latin{et~al.}(2022)Khan, Sprockel, Macmillan, Alting, Kharal,
  Boakye-Ansah, and Haase]{Haase2022}
Khan,~M.~A.; Sprockel,~A.~J.; Macmillan,~K.~A.; Alting,~M.~T.; Kharal,~S.~P.;
  Boakye-Ansah,~S.; Haase,~M.~F. Nanostructured, Fluid-Bicontinuous Gels for
  Continuous-Flow Liquid–Liquid Extraction. \emph{Advanced Materials}
  \textbf{2022}, \emph{34}, 2109547\relax
\mciteBstWouldAddEndPuncttrue
\mciteSetBstMidEndSepPunct{\mcitedefaultmidpunct}
{\mcitedefaultendpunct}{\mcitedefaultseppunct}\relax
\EndOfBibitem
\bibitem[Bruji{\'c} \latin{et~al.}(2003)Bruji{\'c}, Edwards, Grinev, Hopkinson,
  Bruji{\'c}, and Makse]{brujic20033d}
Bruji{\'c},~J.; Edwards,~S.~F.; Grinev,~D.~V.; Hopkinson,~I.; Bruji{\'c},~D.;
  Makse,~H.~A. 3D bulk measurements of the force distribution in a compressed
  emulsion system. \emph{Faraday discussions} \textbf{2003}, \emph{123},
  207--220\relax
\mciteBstWouldAddEndPuncttrue
\mciteSetBstMidEndSepPunct{\mcitedefaultmidpunct}
{\mcitedefaultendpunct}{\mcitedefaultseppunct}\relax
\EndOfBibitem
\bibitem[Fröhlich \latin{et~al.}(2005)Fröhlich, Niedermeier, and
  Luginsland]{frohlich2005_fillerElastomer}
Fröhlich,~J.; Niedermeier,~W.; Luginsland,~H.-D. The effect of filler–filler
  and filler–elastomer interaction on rubber reinforcement. \emph{Composites
  Part A: Applied Science and Manufacturing} \textbf{2005}, \emph{36},
  449--460\relax
\mciteBstWouldAddEndPuncttrue
\mciteSetBstMidEndSepPunct{\mcitedefaultmidpunct}
{\mcitedefaultendpunct}{\mcitedefaultseppunct}\relax
\EndOfBibitem
\bibitem[Gong \latin{et~al.}(2003)Gong, Katsuyama, Kurokawa, and
  Osada]{Gong_DNhydrogels}
Gong,~J.; Katsuyama,~Y.; Kurokawa,~T.; Osada,~Y. Double-Network Hydrogels with
  Extremely High Mechanical Strength. \emph{Advanced Materials} \textbf{2003},
  \emph{15}, 1155--1158\relax
\mciteBstWouldAddEndPuncttrue
\mciteSetBstMidEndSepPunct{\mcitedefaultmidpunct}
{\mcitedefaultendpunct}{\mcitedefaultseppunct}\relax
\EndOfBibitem
\bibitem[Zhang \latin{et~al.}(2022)Zhang, Kim, Hassan, and Suo]{Zhang2022}
Zhang,~G.; Kim,~J.; Hassan,~S.; Suo,~Z. Self-assembled nanocomposites of high
  water content and load-bearing capacity. \emph{Proceedings of the National
  Academy of Sciences} \textbf{2022}, \emph{119}, e2203962119\relax
\mciteBstWouldAddEndPuncttrue
\mciteSetBstMidEndSepPunct{\mcitedefaultmidpunct}
{\mcitedefaultendpunct}{\mcitedefaultseppunct}\relax
\EndOfBibitem
\bibitem[Siddhant~Kumar and Kochmann(2020)Siddhant~Kumar, and
  Kochmann]{Kumar2022}
Siddhant~Kumar,~L.~Z.,~Stephanie~Tan; Kochmann,~D.~M. Inverse-designed
  spinodoid metamaterials. \emph{npj Computational Materials} \textbf{2020},
  \emph{6}\relax
\mciteBstWouldAddEndPuncttrue
\mciteSetBstMidEndSepPunct{\mcitedefaultmidpunct}
{\mcitedefaultendpunct}{\mcitedefaultseppunct}\relax
\EndOfBibitem
\bibitem[Gibson(2005)]{Gibson2005_cellularSolids}
Gibson,~L.~J. Biomechanics of cellular solids. \emph{Journal of Biomechanics}
  \textbf{2005}, \emph{38}, 377--399\relax
\mciteBstWouldAddEndPuncttrue
\mciteSetBstMidEndSepPunct{\mcitedefaultmidpunct}
{\mcitedefaultendpunct}{\mcitedefaultseppunct}\relax
\EndOfBibitem
\bibitem[Zeng \latin{et~al.}(2018)Zeng, Ribbe, Kim, and
  Hayward]{Zeng2018_anisotBlock}
Zeng,~D.; Ribbe,~A.; Kim,~H.; Hayward,~R.~C. Stress-Induced Orientation of
  Cocontinuous Nanostructures within Randomly End-Linked Copolymer Networks.
  \emph{ACS Macro Letters} \textbf{2018}, \emph{7}, 828--833\relax
\mciteBstWouldAddEndPuncttrue
\mciteSetBstMidEndSepPunct{\mcitedefaultmidpunct}
{\mcitedefaultendpunct}{\mcitedefaultseppunct}\relax
\EndOfBibitem
\bibitem[Jr \latin{et~al.}(2008)Jr, Cheng, Bettinger, Borenstein, Langer, and
  Freed]{Engelmayr2008_cardiacAnis}
Jr,~G. C.~E.; Cheng,~M.; Bettinger,~C.~J.; Borenstein,~J.~T.; Langer,~R.;
  Freed,~L.~E. Accordion-like honeycombs for tissue engineering of cardiac
  anisotropy. \emph{Nature Materials} \textbf{2008}, \emph{7},
  1003–1010--399\relax
\mciteBstWouldAddEndPuncttrue
\mciteSetBstMidEndSepPunct{\mcitedefaultmidpunct}
{\mcitedefaultendpunct}{\mcitedefaultseppunct}\relax
\EndOfBibitem
\bibitem[Yongdae~Shin \latin{et~al.}(2019)Yongdae~Shin, Lee, Berry, Sanders,
  Ronceray, Wingreen, Haataja, and
  Brangwynne]{Shin_2006_CondesatesElasticityNucleous}
Yongdae~Shin,~Y.-C.~C.; Lee,~D. S.~W.; Berry,~J.; Sanders,~D.~W.; Ronceray,~P.;
  Wingreen,~N.~S.; Haataja,~M.; Brangwynne,~C.~P. Liquid Nuclear Condensates
  Mechanically Sense and Restructure the Genome. \emph{Cell} \textbf{2019},
  \emph{175}, 1481--1491\relax
\mciteBstWouldAddEndPuncttrue
\mciteSetBstMidEndSepPunct{\mcitedefaultmidpunct}
{\mcitedefaultendpunct}{\mcitedefaultseppunct}\relax
\EndOfBibitem
\bibitem[Wiegand and Hyman(2020)Wiegand, and Hyman]{wiegand2020drops}
Wiegand,~T.; Hyman,~A.~A. Drops and fibers—how biomolecular condensates and
  cytoskeletal filaments influence each other. \emph{Emerging topics in life
  sciences} \textbf{2020}, \emph{4}, 247\relax
\mciteBstWouldAddEndPuncttrue
\mciteSetBstMidEndSepPunct{\mcitedefaultmidpunct}
{\mcitedefaultendpunct}{\mcitedefaultseppunct}\relax
\EndOfBibitem
\bibitem[Ronceray \latin{et~al.}(2022)Ronceray, Mao, Ko{\v{s}}mrlj, and
  Haataja]{ronceray2022liquid}
Ronceray,~P.; Mao,~S.; Ko{\v{s}}mrlj,~A.; Haataja,~M.~P. Liquid demixing in
  elastic networks: Cavitation, permeation, or size selection?
  \emph{Europhysics Letters} \textbf{2022}, \emph{137}, 67001\relax
\mciteBstWouldAddEndPuncttrue
\mciteSetBstMidEndSepPunct{\mcitedefaultmidpunct}
{\mcitedefaultendpunct}{\mcitedefaultseppunct}\relax
\EndOfBibitem
\bibitem[Peng \latin{et~al.}(2021)Peng, Tomsia, Jiang, Tang, and
  Cheng]{peng2021stiff}
Peng,~J.; Tomsia,~A.~P.; Jiang,~L.; Tang,~B.~Z.; Cheng,~Q. Stiff and tough
  PDMS-MMT layered nanocomposites visualized by AIE luminogens. \emph{Nature
  Communications} \textbf{2021}, \emph{12}, 4539\relax
\mciteBstWouldAddEndPuncttrue
\mciteSetBstMidEndSepPunct{\mcitedefaultmidpunct}
{\mcitedefaultendpunct}{\mcitedefaultseppunct}\relax
\EndOfBibitem
\bibitem[Phillip \latin{et~al.}(2010)Phillip, O’Neill, Rodwogin, Hillmyer,
  and Cussler]{phillip2010self}
Phillip,~W.~A.; O’Neill,~B.; Rodwogin,~M.; Hillmyer,~M.~A.; Cussler,~E.
  Self-assembled block copolymer thin films as water filtration membranes.
  \emph{ACS applied materials \& interfaces} \textbf{2010}, \emph{2},
  847--853\relax
\mciteBstWouldAddEndPuncttrue
\mciteSetBstMidEndSepPunct{\mcitedefaultmidpunct}
{\mcitedefaultendpunct}{\mcitedefaultseppunct}\relax
\EndOfBibitem
\bibitem[Lee \latin{et~al.}(2022)Lee, Han, Lee, Lee, Kim, Jung, Kim, and
  Lee]{Lee2022_batteries}
Lee,~M.~J.; Han,~J.; Lee,~K.; Lee,~Y.~J.; Kim,~B.~G.; Jung,~K.-N.; Kim,~B.~J.;
  Lee,~S.~W. Elastomeric electrolytes for high-energy solid-state lithium
  batteries. \emph{Nature} \textbf{2022}, \emph{601}, 217–222\relax
\mciteBstWouldAddEndPuncttrue
\mciteSetBstMidEndSepPunct{\mcitedefaultmidpunct}
{\mcitedefaultendpunct}{\mcitedefaultseppunct}\relax
\EndOfBibitem
\bibitem[Style \latin{et~al.}(2015)Style, Boltyanskiy, Allen, Jensen, Foote,
  Wettlaufer, and Dufresne]{Style2015}
Style,~R.~W.; Boltyanskiy,~R.; Allen,~B.; Jensen,~K.~E.; Foote,~H.~P.;
  Wettlaufer,~J.~S.; Dufresne,~E.~R. Stiffening solids with liquid inclusions.
  \emph{Nature Physics} \textbf{2015}, \emph{11}\relax
\mciteBstWouldAddEndPuncttrue
\mciteSetBstMidEndSepPunct{\mcitedefaultmidpunct}
{\mcitedefaultendpunct}{\mcitedefaultseppunct}\relax
\EndOfBibitem
\bibitem[Landau and Ginzburg(1950)Landau, and Ginzburg]{Landauginzburg}
Landau,~L.~D.; Ginzburg,~V.~L. On the theory of superconductivity. \emph{Zh.
  Eksp. Teor. Fiz.} \textbf{1950}, \emph{20}, 1064\relax
\mciteBstWouldAddEndPuncttrue
\mciteSetBstMidEndSepPunct{\mcitedefaultmidpunct}
{\mcitedefaultendpunct}{\mcitedefaultseppunct}\relax
\EndOfBibitem
\bibitem[K{\"o}nig \latin{et~al.}(2021)K{\"o}nig, Ronsin, and
  Harting]{konig2021two}
K{\"o}nig,~B.; Ronsin,~O.~J.; Harting,~J. Two-dimensional Cahn--Hilliard
  simulations for coarsening kinetics of spinodal decomposition in binary
  mixtures. \emph{Physical Chemistry Chemical Physics} \textbf{2021},
  \emph{23}, 24823--24833\relax
\mciteBstWouldAddEndPuncttrue
\mciteSetBstMidEndSepPunct{\mcitedefaultmidpunct}
{\mcitedefaultendpunct}{\mcitedefaultseppunct}\relax
\EndOfBibitem
\end{mcitethebibliography}

\providecommand{\latin}[1]{#1}
\makeatletter
\providecommand{\doi}
  {\begingroup\let\do\@makeother\dospecials
  \catcode`\{=1 \catcode`\}=2 \doi@aux}
\providecommand{\doi@aux}[1]{\endgroup\texttt{#1}}
\makeatother
\providecommand*\mcitethebibliography{\thebibliography}
\csname @ifundefined\endcsname{endmcitethebibliography}
  {\let\endmcitethebibliography\endthebibliography}{}

\end{document}